\documentstyle[12pt]{article}

        \def\be{\begin{equation}}
       \def\bea{\begin{eqnarray}}

        \def\o{\over}

        \def\l{\label}

        \def\m{\mu}
        \def\n{\nu}

        \def\ee{\end{equation}}
        \def\eea{\end{eqnarray}}

        \def\re{\ref}
        
\begin{document}
\begin{titlepage}
\vspace*{5mm}
\begin{center}
{\Large \bf $Q{\bar Q}$ potential in the Schwinger model
on curved space--time}
\vskip 1cm
{\bf H. Mohseni Sadjadi \footnote {e-mail:amohseni@khayam.ut.ac.ir},
Kh. Saaidi \footnote {e-mail:lkhled@molavi.ut.ac.ir}}
\vskip 1cm
{\it Physics Department, University of Tehran, North Kargar,} \\
{\it Tehran, Iran }\\

\end{center}
\vskip 2cm
\begin{abstract}
We study the confining and screening aspects of the Schwinger model on
curved static backgrounds.
\end{abstract}
\begin{center}
\end{center}
\end{titlepage}
\newpage
\section{ Introduction }

The Schwinger model or two dimensional quantum electrodynamics is one of
the exactly soluble models of quantum field theory \cite{schwinger}.
One of the motivations for studying this model is that it provides a
good laboratory in order to illustrate some important effects
which are present also in four dimensions like screening, confinement,
anomalies and chiral symmetry breaking.

The study of the Schwinger model on a non--dynamical curved background
can be viewed as a first step to study the model in the presence of
quantum gravity. Although the kinetic term of the gauge fields spoils
the conformal invariance, there are many applications
in string theory and quantum gravity coupled to non--conformal matter.

In two dimensions, Maxwell field theory is confining. This is due
to linear rise of the Coulomb potential in one--space dimension.
In the presence of massless dynamical fermions,
via a two dimensional peculiar Higgs phenomenon induced
by vacuum polarization, the gauge field becomes massive and
the Coulomb force is replaced by a finite range force. In the case where
the fermions are massive the model no longer admits an exact solution
but a semiclassical analysis reveals a linear potential between
two opposite external charges which are widely separated.

An important question is how a nontrivial structure of space--time
modifies the above results. In \cite{gass}, by comparison the roles of
temperature and the curvature, it has been suggested that the curvature
may change the confining behavior of the system. In \cite{ghosh}, it has
been shown that, for a particular two dimensional black hole
the massless Schwinger model stills in screening phase.
As we will see this is only true for asymptotically
flat spaces and for example it is not valid for spaces with constant
curvature (like de Sitter spaces).

In this paper we obtain the interaction energy of two static charges
in the Schwinger model on static classical curved background and
we extend our results to non--abelian situation.
Our method is similar to \cite{linet}, in which the self--energy
of a point static charge is computed in four dimensional curved background.
Then using the bosonization rules and by employing the covariant
point splitting method in the regularization of composite operators,
we determine the chiral condensate and discuss briefly the confining
aspects of massive Schwinger model on the curved space--time.

{\section { Massless Schwinger model on static curved space--time }}

Since all two dimensional spaces are conformally flat, the most general
static two dimensional space--time can be described geometrically by
the metric
\be \l{1}
ds^2=g_{\m \n}dx^\m dx^\n=\sqrt{g}(dt^2-dx^2),
\ee
where the conformal factor $\sqrt g$, is only a function of
spatial coordinate. On this space, the Schwinger model is defined
by the action \cite{barcelos}
\be \l{2}
S=\int \sqrt g[-{1\o 4} F_{\m \n}F^{\m \n} +i{\bar \psi}\gamma^{\m}
(\partial_{\m} -ieA_{\m})\psi].
\ee
$\partial_{\mu}$ is the covariant derivative, including the spin connection, acting on fermions in the
curved space--time. The gamma matrices in curved space--time are related to those in Minkowski
flat space $\gamma^a$ by $\gamma^\m=e^\m_a\gamma^a$, where the zweibeins
are defined through
\be \l{3}
g_{\m \n}=e^a_\m e^b_\n \eta_{ab}, \ \ \ \ \  g^{\m \n}
=e^\m_ae^\n_b \eta^{ab}.
\ee
Here $\eta^{ab}=\eta_{ab}=diag(1,-1)$. $F_{\m \n}$ is the electromagnetic
field strength and $e$ is the charge of dynamical fermions. The gravitational
field $g_{\m \n}$ is assumed to be a classical
background. The bosonized version of (\re{2}) is \cite{barcelos}
\be \l{4}
S_B=\int [{1\o {2\sqrt g}} F^2 +{e\over \sqrt \pi}F\phi +{1\o 2}
{(\partial_\m \phi)}^2 ]d^2x,
\ee
where $F={1\o 2}{\hat \epsilon}^{\m \n}F_{\m \n}$ and
${\hat \epsilon}^{01}={\hat \epsilon}_{01}=-{\hat \epsilon}_{10}=1$.
By integrating over the field $\phi $ (or equivalently over $\psi $
in (\re{2})), we arrive at the following effective action
for the gauge fields
\be \l{5}
L_{eff.}={1\o 2\sqrt g}F^2 +{\m^2\o 2}F{1\o {\partial^2}}F,
\ee
where $\m ={e\o {\sqrt \pi}}$. We restrict ourselves to static fields
and choose the Coulomb gauge $A_1=0$. By these assumptions (\re{5})
reduces to
\be \l{6}
L_{eff.}= {1\o 2\sqrt g}({dA_0\over dx})^2 +{\m^2\o 2}A_0^2.
\ee
This Lagrangian shows that like the flat case, the photon has become massive
via a peculiar two dimensional version of Higgs phenomenon,
induced by vacuum polarization. A consequence of this effect
on the flat space, is replacement of the Coulomb force by a finite
range force. To study
this effect (screening) on curved space--time, we introduce two static (with
respect to the coordinates (\re{1})) opposite charges in the
system \cite{gross}. The contravariant
current vector of a particle of charge $e$, which reduces to the current
$J^\alpha =e\int \delta^2(x- x_e)dx^\alpha_e$ in the flat space, is
\cite{weinberg}
\be \l{7}
J^\m ={e\o \sqrt g}\int \delta^2(x-x_e)dx^\mu _e,
\ee
where $dx^\m_e$ is along the  trajectory of the particle.
The static current describing two opposites charges $-e'$ and $e'$ located at
$x=a$ and $x=b$, is then described by
\be \l{8}
J^0(x) ={e'\o{\sqrt g}}(\delta (x-b) -\delta (x-a)), \ \ \ \  J^1=0.
\ee
This current is covariantly conserved
\be \l{9}
{1\o \sqrt g}\partial_\m \sqrt g J^\m=0,
\ee
and adds the interaction term $\int {\sqrt g}J^0A_0d^2x$ to the action.
The equation of motion for the gauge field $A_0$ is
\be \l{10}
{d\o dx}{1\o {\sqrt g}}{dA_0\o dx}- \m^2 A_0= e'(\delta (x-b)-\delta (x-a)).
\ee
The Green function of the self adjoint operator ${d\o dx}
{1\o \sqrt g}{d\o dx}-\m^2$ satisfies
\be \l{11}
({d\o dx}{1\o \sqrt g}{d\o dx}- \m^2)G(x,x')=\delta (x,x').
\ee
In terms of this Green function the solution of eq.(\re{10}) can be
expressed as
\be \l{12}
A_0(x)=\int J^0(x')G(x,x'){\sqrt g(x')}dx'.
\ee
By substituting (\re{12}) back into the action, the energy of external
charges is obtained
\be \l{13}
E=\int T^0_0dx=-\int L_{eff.}dx=-{{e'^2}\o 2}[G(a,a)+G(b,b)-2G(a,b)].
\ee
$T^{\m \n}$ is the energy momentum tensor, thus $T^0_0$ is the energy
density measured by an observer whose velocity  $u^\m =(g^{-1/4},0)$
is parallel to the direction of the timelike killing vector of
the space--time. ${-e'^2\o 2}[G(x,x)\equiv \lim_{y\rightarrow x}G(x,y)]$
is the change of the energy when a static point charge located at $x$
is added to the system and $e'^2 G(a,b)$ is the mutual interaction potential
between external charges.

On the flat space $G(a,b)={-1\o 2\m}e^{-\m|b-a|}$, thus $G(x,x)$ is a
constant
and for largely separated charges the energy is equal to $e'^2\o 2\m $
(using (\re{13}) one can show that this is also valid for infinite asymptotically
flat space--time).
Physically this is due to the screening of external charges by dynamical
fermions.

On the curved space the behavior of the energy depends on the form of the
Green function.
In order to show the effect of the metric on the energy, let us
consider the Schwinger model on some special curved spaces. Firstly we
consider a two dimensional space with small curvature with respect to
the coupling $\mu $. Writing the homogeneous solution of the
eq.(\re{11}) as $G_h={d\Phi \o dx}$,
one can see
that $\Phi$ satisfies the equation
\be \l{16}
{d^2\Phi \o dx^2}- \m^2\sqrt g\Phi =0.
\ee
This is the equation of motion for a static
massive scalar field on the conformally flat space--time ({1}).
Equation (\re{16}) possesses formal WKB solution
\be  \l{17}
\Phi ={c\o W^{1\o 2}}exp[\int^x W(x')dx'],
\ee
where $c$ is a constant. $W$ satisfies the nonlinear equation
\be \l{18}
W^2=\m^2\sqrt g+{1\o 2}({W''\o W} -{3\o 2}{W'^2 \o W^2}).
\ee
If the metric is slowly varying with respect to the coupling $\m$,
then $({g^{1/4}})'\ll \m \sqrt g$,
so the zeroth order approximation is to substitute $W^0=\pm \m g^{1/4}$ into
(\re{17}). The next iteration
of eq.(\re{18}) yields
\be \l{19}
(W^{(2)})^2=\mu^2\sqrt g +{1\o 2}({{g^{1\o 4}}''\o g^{1\o 4}} -{3\o 2}
{[(g^{1\o 4})']^{2}\o g^{1\o 2}}).
\ee
The higher order can be obtained by continuing the iteration of eq.(\re{18}).
At the zeroth order approximation
(neglecting the derivative of
the metric), the Green function (\re{11}) is obtained as
\be \l{20}
G(x,x')={-1\o{2\m}}g^{1\o 8}(x)g^{1\o 8}(x')e^{-\m |\int_{x}^{x'}
g^{1\o 4}(y)dy|}.
\ee
One can directly check that the Green function (\re{20}) satisfies
the eq.(\re{11}), on condition that $({g^{1/4}})'\ll \mu \sqrt g$ and
$R \ll \mu^2$, where $R$ is the scalar curvature of the space.
The energy is then
\be \l{21}
E={e'^2\o 4\m }[g^{1\o 4}(a)+g^{1\o 4}(b) -2g^{1\o 8}(a)g^{1\o 8}(b)
e^{-\m |\int^{b}_{a} g^{1\o 4}(x)dx|}].
\ee
Setting $g=1$, we recover the well--known result
\be \l{22}
E_{flat.}= {{e'}^2\o 2\m}(1-e^{-\m |b-a|}).
\ee
On the curved space $|a-b|$ has been replaced by the geodesic distance
$d_c=|\int ^b_a  g^{1\o 4}dx|$. In the limit $d_{c}\rightarrow \infty$,
the energy tends to 
\be \l{23}
E={{e'}^2\o 4\mu}[g^{1\o 4}(a)+g^{1\o 4}(b)],
\ee
which in contrast to the flat case is not a constant and depends on
the value of the metric at the position of external charges.
Hence the presence of the metric modifies
the confining behavior of the system.
As an illustration consider the covering of two dimensional de Sitter space
\cite{yokoi}
\be \l{24}
ds^2={1\o {\rm cos}^2({x\o \rho})}(dt^2-dx^2),\ \ \ -\infty <t<\infty;
{-\pi \o 2}<{x\o \rho}<{\pi\o 2}
\ee
with scalar curvature $R={2\o \rho^2}$, where ${2\o \rho^2}\ll \m^2 $.
The energy of external charges is
\be \l{25}
E={{e'}^2\o 4\m}[{1\o {\rm cos}({a\o \rho})}+{1\o {\rm cos}({b\o \rho})}-
{2\o {\rm cos}^{1\o 2}({a\o \rho}){\rm cos}^{1\o 2}({b\o \rho})}
{\rm exp}(-\m |
\int^b_a{1\o {\rm cos}({x\o \rho})}dx|)].
\ee
If one of the test charges is located at ${x\o \rho}\cong {\pm{\pi \o 2}}$
we have $E\gg {{e'}^2\o {2\m}}$, unless $a\cong b$. This shows that
in  small neighbourhood of $x=\pm {\rho\pi \o 2}$, the system approaches
to the confining phase. When $R=0$ (flat space) this effect disappears.

In the  above we have assumed  that the coupling $\mu$ is much greater
than the scalar curvature of the space. Let us now study the confining nature
of the massless Schwinger model on a space--time with an arbitrary curvature.

Consider the following de Sitter space described by the metric (in
Schwarzschild coordinates)
\be \l{26}
ds^2={r^2\o \lambda^2}dt^2-{\lambda^2\o r^2}dr^2,\ \ \ \ \ r>0.
\ee
This is one of the solutions of the equation $R={2\o \lambda^2}$
obtained by varying the Jackiw--Teitelboim action
\be \l{27}
S=\int d^2x{\sqrt g}\Phi (R-{2\o \lambda^2}),
\ee
with respect to the field $\Phi$.
In the coordinate $(x,t)$ where, $dx^2={\lambda^4\o r^4}dr^2$, $x>0$,
the metric (\re{26}) takes the conformally flat shape
\be \l{28}
ds^2={r^2\o \lambda^2}(dt^2-dx^2).
\ee
The symmetric Green function (\re{11}) in terms of $r$ coordinate is
\be \l{29}
G(r,r')=-{1\o {2l-1}}{r_{<}^{l}\o {r_{>}^{l-1}}},
\ee
where $l={{1+{\sqrt {1+4\m ^2\lambda^2}}}\o 2}$.
This Green function satisfies Dirichlet boundary condition at $r=0$ and
at $x=0$. The energy of external charges is
\be \l{30}
E={e'^2\o 2({2l-1})}(r_a +r_b -2{r_a^l\o r_b^{l-1}}),
\ee
where $r_a=r(a)$ and $r_b=r(b)$ and $r_{b}>r_{a}$. The distance of
separation of static charges in
terms of coordinate $r$ is $d=\lambda ln {r_b\o r_a}$.
In the limit
${r_{b}} \rightarrow \infty$ ($d\rightarrow \infty $),
the energy increases infinitely, therefore for
${1\o \lambda}\gg \mu$
the system is strongly in confining phase.
  
These results can be extended to non--abelian cases. The action of massless
multiflavor $QCD_2$, with fermions in the fundamental representation of
$SU(N_c)$, in the presence of external current $J^{\m }$, is
\be \l{31}
S=\int d^2x\sqrt{g}{\rm tr}({1\o g{e_c}^2}F^2+i{\bar \psi}{\gamma^{\m}}D_{\m}\psi
+J^{\mu }A_{\m}),
\ee
where $\psi =\psi^i_j$, $i=1\cdots N_c; j=1\cdots N_f$ and $N_f (N_c)$ is
the number of flavors (colors). $F={1\o 2}{\hat \epsilon}^{\m \n}
(\partial_{\m}A_{\n}-\partial_{\n}A_{\m}+i[A_{\m},A_{\n}])$ and $D_{\mu}
=\partial_{\m} +iA_{\m}$.

In \cite{fri}, it has been shown that in
the bosonized version of the action (\re{31}), the flavor degrees of
freedom can be separated from the color ones and the colored sector takes
the form
\bea \l{32}
S&=&N_f \{ S_{WZW}(h)- \int d^2x[{\rm tr}{-1\o {N_fe^2_c}\sqrt{g}}F^2 -
{\sqrt{g}\over N_f}{\rm tr}(J^{+}A_{+}+ J^{-}A_{-})\nonumber \\
& \;\;\;\;& +{1\o 2\pi }{\rm tr}(ih^{-1}
\partial_{+}hA_{-} +ih{\partial_{-}}h^{-1}A_{+} + A_{+}hA_{-}h^{-1}-
A_{+}A_{-})] \}.
\eea
$S_{WZW}(h)$ is the action of $WZW$ theory with $h\in SU(N_c)$.
The light cone component of a vector $A$ is defined by
$A_{\pm}={{A_0 \pm A_1}\o \sqrt 2}$.
$A_{\pm}$ take their values in the algebra of $SU(N_c)$
whose generators $T^{i}$ are normalized as
${\rm tr}T^{i}T^{k}={1\o 2}\delta^{ik}$.
On the conformally flat curved space--time (\re{1}), the bosonic part
of the action (\re{32}), is the same as the flat case \cite{ivanov}.
Only the kinetic terms of the gauge fields destroys this invariance.
At large $N_f$ limit the system is classical and
an investigation of the equations of motions, captures the quantum
behavior of the theory. Using the equation of motion for the matter
and gauge fields in the gauge $A_{-}=0$, one can obtain \cite{arm}
\be \l{33}
\partial_{+} {1\o \sqrt g}\partial_{-} A_{+}
+{i\o \sqrt g}[\partial_{-}A_{+},A_{+}] + {e_c^2N_f\o 4\pi}A_{+}
=-{e_c^2\over 2}\sqrt{g} J^{-}=-{e_c^2\over 2}J_{+}.
\ee
We assume that the external current is composed of two static
opposite charges $\pm e'$ located
at the points $a$ and $b$, and points in one of the $N_c^2-1$ directions
of $SU(N_c)$, which we denote $j$.
\be \l{34}
{J^{-}}={J^{+}}={1\o \sqrt{g}}e'T^i
 \delta^{j,i}[\delta (x-b) -\delta (x-a)]
\ee
The abelian static solution of the equation (\re{33}) is
\be \l{35}
A_{+}(x)=\int e_{c}^2G(x,x')J_{+}(x')dx',
\ee
where the static Green function $G(x,x')$, satisfies the equation (\re{11}),
with ${\m ^2}={e_c^2N_f\o 2\pi}$.
The interaction energy is \cite{{gross},{arm}}
\be \l{36}
E={-1\o 2}\int^b_a\sqrt {g}{\rm tr}J^{+}(x)A_{+}(x)dx={-1\o 2}{e'}^2e_c^2{\rm tr}[(T^j)^2]
[G(a,a)+G(b,b)-2G(a,b)],
\ee
which implies the same confining behavior for the system as in the abelian
situation.

\section {Quenched Schwinger model ($\m =0$)}

On the flat space--time and in the absence of dynamical fermions $\mu=0$,
the vacuum polarization is switched off and the screening effect is replaced
by confinement. One way to study this confining behavior is
the computation of the vacuum expectation value of the Wilson loop $C$:
\be \l{37}
<W[C]>=\int DA{\rm exp}(-S+ ie'\oint_{C}dx^{\m }A_{\m }),
\ee
where $C$, is a closed loop on the space--time. When $<W[C]>= \beta
{\rm exp}(\alpha A[C])$, where $\alpha <0$ and $\beta $ are two constants
and $A[C]$ is the area enclosed by $C$, the system is in confining phase.
This means that the potential at a point of the loop,
due to presence of a charge at other points, is proportional to the distance.
As we will see on the static curved space, this behavior is complicated
by the presence of the metric.
In this part, we consider metrics with Euclidean signature obtained
from (\re{1}) by analytical continuation, that is
$ds^2={\sqrt g}(x) (dx^2 +dt^2)$. On this space the geodesic distance
between the points $(a,t)$ and $(b,t)$ is $d=\int_{a}^{b} g^{1\o 4} \, dx $.
Our Wilson loop is a rectangle characterized by points $(-{T\o 2}, a)$,
$(-{T\o 2}, b)$, $({T\o 2}, a)$ and $({T\o 2}, b)$, in this section we
assume $b>a$. We have
\be \l{38}
\oint_{C}dx_{\m}A^{\m}=\int \sqrt {g}A_{\m}I^{\m}d^2x,
\ee
where
\be \l{39}
I^{\m}(x)=\oint_{C} dz^{\m}{\delta^2(x,z)\o \sqrt g}
\ee
(compare with definition (\re{8})).
In the Coulomb gauge $A_1=0$, the Wilson loop expectation value is
\be \l{40}
<W[C]>={{\int DA_0{\rm exp}\int[ie'I_0A_0-{1\o 2\sqrt g}(\partial_1 A_0)^2]
d^2x}\o {\int DA_0{\rm exp}\int -[{1\o 2\sqrt g}(\partial_1 A_0)^2]d^2x}}.
\ee
The functional is saturated by $A_0$ satisfying the following equation
\be \l{41}
{d\o dx}{1\o \sqrt g} {dA_0\o dx}=-ie'I_0,
\ee
with solution
\be \l{42}
A_0(x,t)=-ie'\int I_0({x',t'})G((t,x),(t',x'))dt'dx',
\ee
where
\be \l{43}
G={1\o 2}\delta (t,t')|\int^x_{x'}g^{1\o 2}(y)dy|.
\ee
The expectation value of the Wilson loop is then
\be \l{44}
<W>={\rm exp}[{e'^2\o 2}\oint_C dt\oint_C dt' G((x,t),(x',t'))]
={\rm exp}[{-{{e'}^2}\o 2}T\int^b_a\sqrt g(y)dy].
\ee
This result can be extended to non--abelian gauge fields. By considering
that in the Coulomb gauge there is no ghost contribution we are
left with
\be \l{45}
<W>={{\rm tr}P\int DA_0{\rm exp}(ie'\oint_C A_0dx^{0}){\rm exp}
[{1\o 2}{\rm tr}
\int {-1\o \sqrt{g}}(\partial_{1}A_{0})^2d^2x]\o \int DA_0{{\rm exp}[-{1\o 2}{\rm tr}\int
{1\o \sqrt{g}}(\partial_{1}A_{0})^2d^2x]}}.
\ee
$P$ is the path ordering and $A=A^iT^i$, where $T^i$ are the generators
of the Lie group under consideration. Completing the square in the functional integral
we find \cite{abda}.
\be \l{46}
<W>={\rm tr}P{\rm exp}[{{e'}^2\o 2}\oint_C \oint_C T^i_{t}G((x,t),(x',t'))T^i_{t'}
dtdt'].
\ee
Using the locality of the Green function in $t$ and $t'$, the group indices
are unimportant and the $P$ symbol can be deleted
\be \l{47}
<W>={\rm dim}R{\rm exp}[{-{e'}^2\o 2{\rm dim}R}C_2(R)T\int_a^b \sqrt gdx].
\ee
$C_2(R)={\rm tr}T^iT^i$ is the quadratic Casimir and $R$ is the representation of
the gauge group. $dim R$ is released from taking the trace in (\re{46}).
The energy of external charges is
\be \l{48}
E={-{1\o T}}ln<W>.
\ee
On the flat space the area behavior of the Wilson loop implies a linear
potential between external charges. This lies on the fact that the area
of the Wilson loop is $(b-a)T$. On the curved space the area
$T\int^b_a \sqrt g dx$ is not proportional to the geodesic distance of
charges which is $d=\int_a^b g^{1\o 4}(y)dy$. When the area grows infinitely
in the limit $d\rightarrow \infty$, the system is in confining phase.
As an example consider the two dimensional black hole \cite{ghosh}
\be \l{49}
ds^2=(1+{M\o \lambda}e^{-\lambda x})^{-1}(dt^2-dx^2),\ \ \ \ -\infty<x<\infty,
\ee
where $\lambda^{-1}$ is a length scale and $M$ is the mass of the black hole.
The energy of charges is proportional to
the area $A=T\int^b_a (1+{M\o \lambda}e^{-\lambda x})^{-1}dx$, which grows
infinitely when
$(d=\int^b_a(1+{M\o \lambda}e^{-\lambda x})^{-{1\o 2}}dx)\rightarrow \infty$.
Therefore the system is in confining phase.
There are also some space--times where the area behavior
of the Wilson loop is not a sign of confinement. For example consider
the area of the Wilson loop on the
anti--de Sitter space (known as the Poincare half plane)
$ds^2={1\o x^2}(dt^2+dx^2),\ \ \ x>0$ :
$A=T({1\o a}-{1\o b})$ which tends to a finite value
when $ b \rightarrow \infty $  (or $d={\rm ln}{b\o a} \rightarrow \infty $),
signalling the screening phase. This result is in agreement with 
\cite{callan}, in which using the proportionality of
the perimeter and the area of a large Wilson loop on the Poincare
half plane, it has been claimed that the area behavior of
the Wilson loop is not an appropriate criterion of confinement.

\section {Chiral condensate and bosonization}

The fermionic condensate is one important characteristic
of the vacuum state
in the Schwinger model and is an order
parameter for the chiral symmetry
breaking. The curvature dependence of the chiral condensate
can be obtained from the bosonization rule \cite{eboli}
\be \l{50}
{\bar \psi}\psi =\Sigma(x,t) N_{\m}{\rm cos}(2\sqrt \pi \phi).
\ee
$\Sigma (x,t)$ is field independent and depends on the normal ordering
(with respect to the mass $\m $) in defining the composite operator.

In order to determine $\Sigma $, we must compute the correlator
\cite{smilga} $<{\bar \psi}\psi(P_{1}){\bar \psi} \psi (P_{2})>$,
where $P_1(x_1,t_1)$ and $P_2(x_2,t_2)$ are two points on the space--time.
Like the previous part we consider the Schwinger model on the space
$ds^2=\sqrt g(dx^2+dt^2)$, which is obtained from (\re{1}),
by analytical continuation of the time coordinate.
By performing the (decoupling) change of variables \cite{naon}
\be \l{51}
\psi =e^{e\gamma _5 \varphi}\lambda , \ \ \ \  {\bar \psi}
= {\bar \lambda}e^{e\gamma_5 \varphi},
\ee
$$
A_{\m}={\hat \epsilon }_{\m \n}{\partial _\nu}\varphi,
$$
the action of the Schwinger model
becomes
\be \l{52}
S=\int \sqrt g[-i{\bar \lambda}\gamma^\m \partial_{\m}\lambda +
{1\o 2}\varphi \Delta
(\Delta -{\mu}^2)\varphi ]d^2x.
\ee
$\Delta $ is Laplace-Beltrami
operator $\Delta ={1\o \sqrt g}\partial_\m
g^{\m \n}\sqrt g\partial _\n $. For $g_{\m \n}=\sqrt g diag(1,1)$, this
operator is $g^{\m \n}\partial_\m \partial_\n $.
The term ${\m ^2\o 2}{\sqrt g}\varphi \Delta \varphi $ is related to the
chiral anomaly and is obtained from the Jacobian of the transformation
(\re{51}). The fermionic fields $({\bar \lambda},\lambda )$ are free,
hence the fermionic part of the action is invariant under Weyl transformation \cite{davies}
$$
g_{\m \n} \rightarrow {\bar g}_{\m \n}={\Omega^2}g_{\m \n }
$$
\be \l{53}
\lambda \rightarrow \xi =\Omega^{-{1\o 2}}\lambda,\ \ \ {\bar \lambda}
\rightarrow {\bar \xi}=\Omega^{-{1\o 2}}{\bar \lambda}.
\ee
By taking $\Omega =g^{-{1\o 4}}$, (\re{52}) becomes
\be \l{54}
S=\int [-i{\bar \xi}\gamma^a \partial_a \xi +{\sqrt{g} \o 2}\varphi \Delta
(\Delta - {\mu}^2)\varphi ]d^2x.
\ee
The Jacobian of the transformation (\re{53}), which is related to the
conformal anomaly, depends only on the metric and in our case where the
metric is not quantized, has been suppressed from the action.
Using the above results we deduce
\be \l{55}
<{\bar \psi} {\psi(P_1)}{\bar \psi}\psi (P_2)>={<{\bar \xi}\xi (P_1)
{\bar \xi}\xi (P_2)>\o g^{1\o 4}(P_{1})g^{1\o 4}(P_2)}{\rm exp}
\{ 2e^2[K(P_1,P_1)+ K(P_2,P_2) -2K(P_1,P_2)]\},
\ee
where
\be \l{56}
K(P,P)={\pi \o e^2}[G_0(P,P)-G_{\m}(P,P)],
\ee
and $G_{\m}$ is the Green function of the massive scalar field $\phi $,
computed from the Lagrangian
\be \l{57}
L={-{1\o 2}}\phi \partial^2 \phi +{1\o 2}\sqrt g {\m^2}\phi^2.
\ee
For $\m =0$, $G_{\m}(P_1,P_2)$ is the same as the massless scalar
Green function on the flat space
$G_0(P_1,P_2)=-{1\o 2\pi}{\rm ln}|P_1-P_2|$ \cite{davies}.
By considering that
the free fermionic correlator is $<{\bar \xi}\xi (P_1){\bar \xi}\xi (P_2)>=
{1\o 2\pi ^2|P_1 -P_2|^2}$, we get
\be \l{58}
\lim _{P_2\rightarrow P_1}<{\bar \psi}\psi (P_1){\bar \psi}\psi (P_2)>=
{1\o g^{1\o 2}(P_1)}\lim_{P_2\rightarrow P_1}{1\o 2\pi^2 |P_1-P_2|^2}.
\ee
On the other hand, from (\re{50}), we have
\be \l{59}
<{\bar \psi}\psi (P_{1}){\bar \psi}\psi(P_2)>=\Sigma (P_{1})\Sigma (P_2)<N_{\m }
{\rm cos}[2\sqrt \pi \phi (P_1)]N_{\m}{\rm cos}[2\sqrt \pi \phi (P_2)]>.
\ee
In order to regularize the ultraviolet divergence of the composite operators,
we employ the covariant point splitting method
\be \l{60}
G_{\m}^{\it reg.}(P,P)=<{\phi^2(P)}>^{\it reg.}=\lim_{P'\rightarrow  P}[G_{\m}
(P,P')- G_{\m}^{DS}(P,P')],
\ee
where $G_{\m}^{DS}(P,P')$ is a point splitting counterterm needed to
regularize $G(P,P)$. Denoting by $\epsilon $ the one half of the proper
distance between $P$ and $P'$, $G^{DS}$ has
been obtained as \cite{bunch}
\be \l{61}
G_{\m}^{DS}(P,P)=-{1\o 4\pi}[2\gamma
+{\rm ln}({\m^2}{\epsilon^2})-{R \o {6\m^2}}]+
O(\epsilon^2).
\ee
$\gamma $ is the Euler constant and $R$ is the scalar curvature of the space.
On the flat space the relation (\ref{60}) is reduced to the normal ordering prescription
which kills out the loops. From (\re{59}) we obtain
\be \l{62}
\lim_{P_{2}\rightarrow P_{1}}<{\bar \psi}\psi (P_1) {\bar \psi}\psi (P_2)>
={\Sigma^2(P_1)\o 2}{\rm exp}[-4\pi G_{\m}^{reg.}
(P_1,P_1)]\lim _{P_2\rightarrow P_1}{\rm exp}[4\pi
G_{\m}(P_1,P_2)].
\ee
To derive this equation we have used $\exp[-G_{\mu}(P,P)]=0$ \cite{davies}.
Comparing (\re{62}) and (\re{58}) results
\be \l{63}
\Sigma^2(P_1)=\lim_{P_2\rightarrow P_1}{g^{-{1\o 2}}(P_1)\o {\pi^2}|P_1-P_2|^2}{\rm exp}
[-4\pi G_\m^{DS}(P_1,P_2)],
\ee
which yields
\be \l{64}
 {\bar \psi}\psi(P_1)={-g^{-{1\o 4}}(P_1)\o {\pi}}{\rm exp}
 \{-2\pi [G_{\m}^{DS}(P_1,P_1)-G_0(P_1,P_1)]\}
 N_{\m}{\rm cos}[2\sqrt \pi \phi(P_1)].
\ee
Therefore the chiral condensate is
\be \l{65}
<{\bar \psi}\psi(P)>={-1\o \pi}g^{-{1\o 4}}(P){\rm exp}[2e^2K(P,P)],
\ee
The Green function $G_\mu $ is known only for very particular curved space.
Here we only need the short distance expansion of $G_{\m}$. One can use the
adiabatic expansion method, to get \cite{chris}
\be \l{66}
G_{\m}(P,P)=\lim_{P'\rightarrow P}{1\o 4i}\sum_{j=0}^{\infty}a_j(P',P)
({-\partial \o \partial{\m^2}})^j H^2_0(\mu s).
\ee
$H_0^2$ is the Hankel function of the second kind and $s$,
is the geodesic distance of points $P$ and $P'$.
Substituting (\re{66}) into (\re{65}) yields
\be \l{67}
<{\bar \psi}\psi(P)>_{R}=<{\bar \psi}\psi>_{R=0}exp[{-{1\o 2}}e^2
\sum_{j=1} a_j(P,P){(j-1)!\o {\m^{2j}}}],
\ee
where $a_i$ are the Seeley DeWitt coefficients, which are computed up to
$i=5$ in \cite{avramidi} and
$<{\bar \psi}\psi>_{R=0}=-{{\rm exp}\gamma \o {2\pi^{3\o 2}}}e$ \cite{jay}.
The result (\re{67}) coincides exactly with the results of ref.\cite{wipf}.

\section {Massive Schwinger model}

The massive Schwinger model, describing electrodynamics of massive
dynamical fermions on the curved space (\re{1}), is defined by the partition
function
\be \l{68}
Z=\int DA_{\m}D{\bar \psi}D\psi \exp[i\int \sqrt g
d^2x(i{\bar \psi}\gamma^{\m}
(\partial_{\m}-ieA_{\m})\psi-m{\bar \psi}\psi-{1\o 4}F^{\m \n}F_{\m \n})].
\ee
$m$ is the mass of dynamical fermions. Introducing the external charges
(\re{8}), adds ${\sqrt g}J^0A_0$ to the action. By performing the change
of variable (\re{53}), the fermionic part of (\re{68}) becomes the same
as the flat case with a position dependent mass term.
On the space (\re{1}), the partition function is
\be \l{69}
Z=\int DA_{\m}D{\bar \xi}D\xi exp\{i\int d^2x[i{\bar \xi}\gamma^a(\partial_a-
ieA_a)\xi-mg^{1\o 4}{\bar \xi}\xi+{1\o 2}{F^2\o \sqrt g}+ J_0A_0]\}
\ee
To eliminate the external current we perform the field rotation \cite{armoni}
$$
\xi \rightarrow e^{{i\o 2}\alpha (x)(1-\gamma^5)}\xi
$$
\be \l{70}
{\bar \xi} \rightarrow {\bar \xi}e^{{-{i\o 2}}\alpha (x)(1+\gamma^5)},
\ee
where $\alpha(x)=-2\pi {e'\o e}[\theta (x-b) -\theta (x-a)]$, $b>a$
and $\theta$
is the step function. The interaction term ${\sqrt g}J^0A_0$,
is cancelled out exactly by ${e\partial_{1}\alpha (x)\o 2\pi}A_{0}$ which is the Jacobian
of the transformation (\re{70}). We assume that external changes are
largely separated such that the spatial part of the space under study is
bounded by
these charges \cite{armoni}. By applying these assumptions in (\re{68})
we arrive at
\be \l{71}
Z=\int DA_{\m}D{\bar \xi}D\xi exp[i\int d^2x({1\o 2\sqrt g}F^2+i{\bar \xi}
\gamma^a(\partial_a-ieA_{a})\xi-mg^{1\o 4}{\bar \xi}
e^{-2\pi i{e'\o e}\gamma^5}\xi)].
\ee
The energy of external charges is $<H>-<H>_0$, where $<H>$ ($<H>_0$), is the
vacuum expectation value of the Hamiltonian in the presence (absence) of
external charges. This change is due to the mass term. By considering
$<H>=\int <T^0_0>dx$, and  parity invariance of the massless part of
(\re{71}), one can obtain the
energy up to the first order of $m$ as \cite{armoni}
\be \l{72}
E=-m\int^b_a \sqrt g <{\bar \psi}\psi >_{m=0}[1-{\rm cos}(2\pi {e'\o e })]dx.
\ee
For a general metric the computation of the energy is complicated by
the presence of the Seeley--Dewitt coefficient in (\re{67}).
For a de Sitter space in which $<{\bar \psi} \psi >$ is a constant \cite{wipf},
the energy is
\be \l{73}
E=-m[1-{\rm cos}(2\pi {e'\o e})]<{\bar \psi}\psi > \int^b_a\sqrt g dx.
\ee
which for ${e'\o e}\notin Z$ has a similar behavior as the quenched Schwinger model.
For ${e'\o e} \in Z$, the energy vanishes. This is due to the screening
of external charges by dynamical charges.
This is the dominant part of the energy for largely separated charges
in first power of $m$ and must be modified by
implying the short range corrections.  We will discuss this elsewhere.

\end{document}